\documentclass[11pt,twoside]{article}

\usepackage{asp2006}
\usepackage{lscape}

\begin{document}
\title{Starburst in the interacting HII galaxy II Zw 40 and in non-interacting HII galaxies}   
\author{Eduardo Telles\altaffilmark{1,2}}  
\altaffiltext{1}{Observat\'orio Nacional, Rua Jos\'e Cristino, 77, Rio de Janeiro,
RJ, 20921-400, Brazil}
\altaffiltext{2}{Astronomy Department, University of Virginia, P.O. Box 400325, Charlottesville, VA 22904-4325, USA}



\begin{abstract} 
In this poster, I summarize the results of our integral field spectroscopic observations of
the nearby prototype of HII galaxies, II Zw 40.  Observations with GMOS-IFU on GEMINI-North
in the optical allowed us to make a detailed kinematic picture of the central starburst,
while SINFONI with adaptive optics on the ESO-VLT gave us a near-IR view of the
interplay between the ISM phases.

Here, I also address the question that not all starbursts require an external trigger such as 
a galaxy-galaxy encounter, as it seems to be the case for a fraction of low luminosity HII galaxies.
We speculate that these may form stars spontaneously like ``pop-corn in a pan''.

\end{abstract}



\section*{Kinematics and the ISM of II Zw40}

We recall our study of the kinematic properties of the ionized gas in the
dominant giant H{\sc ii} region of this well known H{\sc ii} galaxy: II Zw 40
\citep{bor09}. With the 3D spectroscopy we obtained the H$\alpha$ intensity map,
the radial velocity and velocity dispersion maps as well as estimate some
physical conditions in the inner region of the starburst, such as oxygen
abundance (O/H) and electron density. The analysis of a set of kinematics
diagnostic diagrams, such as the intensity $versus$ velocity dispersion
(\textit{I}-$\sigma$), intensity $versus$ radial velocity
(\textit{I}-\textit{V}) and $V$-$\sigma$, for global and individual analysis in
sub-regions of the nebula allowed us to separate the main line broadening
mechanisms responsible for producing a  smooth supersonic integrated line
profile for the giant H{\sc ii} region. The deconvolution of the effects of
stellar winds and possibly SN, i.e. bubbles, shells, revealed regions of
``unperturbed'' motions over the whole extent of the starburst which are still
supersonic ($\sim$ 30 km s$^{-1}$). We interpret these motions as being related
to the underlying gravitational potential and dominant in very young regions
when measured through the integrated aperture over the central starburst region.
Our observations show that the complex structure of the interstellar medium of
this galactic scale star-forming region is very similar to that of nearby
extragalactic giant H{\sc ii} regions in the Local Group galaxies.

The central starburst region in II Zw 40 was also observed with SINFONI integral
field spectroscopy in the near infrared with adaptive optics provided by MACAO
on the VLT \citep{van08}. We  assessed the interplay of the phases of the ISM by
mapping the fluxes, radial velocities and velocity dispersions of Br$\gamma$,
Fe[II] and molecular Hydrogen H$_2$. The radiation emitted by the galaxy is
dominated by a giant HII region which extends over an area of more than 400 pc
in size and is powered by a very young stellar population.  The spatial
distribution and velocity field of different components of the ISM, mostly
through the Bracket series lines, the molecular hydrogen spectrum and [FeII]
tell us  that [FeII] and $H_2$ are mostly photon excited but, while the region
emitting [FeII] is almost coincident with the giant HII region observed in the
lines of atomic H and He, the $H_2$ has a quite different distribution in space
and velocity. The age of the stellar population in the main cluster is such that
no SN should be present yet so that the gas kinematics must be dominated by the
young stars. The starbursting region  seems to be detached geometrically or
dynamically from the large scale morphology of the galaxy. In other words, the
properties of the starburst region does not keep memory of what caused it!

\section*{Are all starbursts triggered by interactions?}

Our study on the morphology of a sample of HII galaxies has revealed that the
they are irregular dwarf galaxies with widespread recent star formation activity
extending over most of the optical images \citep{tmt97}.  But we noted that the more
luminous HII galaxies (type I) show signs of disturbances on their outer
envelopes while the less luminous HII galaxies (type II) are more regular in
their shapes. The most compact, with little or no signs of an extended envelope,
were found among the least luminous galaxies.  However, the position, number and
sizes of the star forming regions or knots were unrelated to the their overall
shapes.  It seems natural to explain the disturbances in type I HII galaxies as
being due to ongoing mergers and advanced interactions between dwarf galaxies.
On the other hand, type II HII galaxies defy our attempts to insert them in the
same scenario.  The starburst regions do not consist of one or two knots but of
ensembles of star forming knots which do not often concentrate in the optically
deduced center of the galaxy \citep[see also][]{lt85}. More recently, we used narrow band surface
photometry to assess the morphology of the star-forming regions as compared to
their true continuum emission \citep{lag07}.  With  higher spatial resolution
observations an even more fragmented view of the starburst regions was revealed,
where different star forming knots were associated with ensembles of star
clusters spread over the extent of the galaxy with no relation with the overall
morphology. Again, it seems that by the morphology, the multiplicity and likely
other properties of the starburst region alone,   we cannot infer its triggering
mechanism.  These dwarf starbursting galaxies cannot be products or debris of
strongly interacting systems because they are not at all associated with giant
galaxies \citep{tt95}. In fact, they seem to populate low density regions and
are not particularly more clustered than the normal field population
\citep{tm00}. Other attempts have been made to search for companions around HII
galaxies, though the results are inconclusive. Even then, one would still need
to explain how smaller galaxies \citep{pus01} or HI clouds \citep{tay95} would
have a significant tidal effect over large distances. A lesson is well learnt
that interactions cause starbursts, but is the inverse question true?

\section*{Galaxies that form star clusters like ``pop corn''} 

\begin{figure}[ht!]
\centering
\plotone{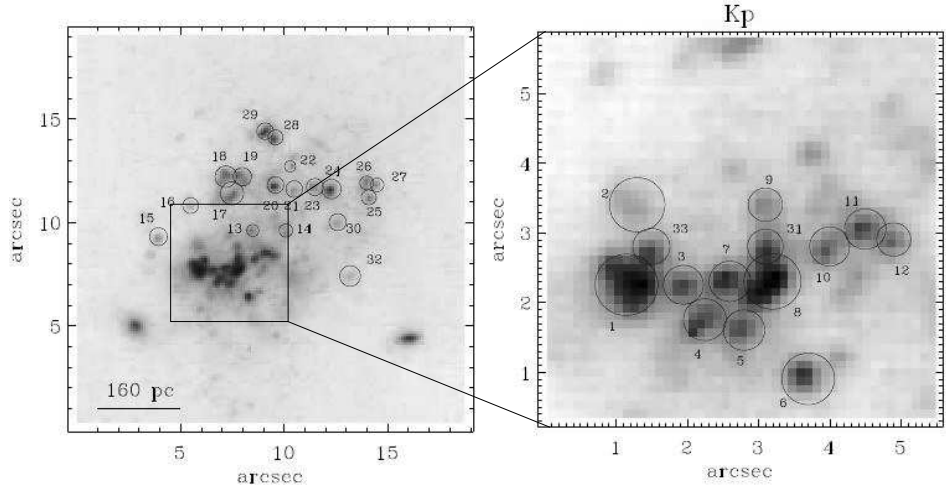}
\caption{Mrk 36 (Haro 4) GEMINI-NIRI K-band image from Lagos etal. (2009, in prep). 
This high spatial resolution image shows unprecedented details and resolved star clusters spread over
the whole extent of this HII galaxy.}
\label{fig1} 

\end{figure}

HII galaxies and Blue Compact Galaxies have mostly identical properties, though
they differ historically by their selection criteria. Despite of their original
compact appearance in photographic plates, they have revealed a more complex
structure when observed with high spatial resolution. Like in the more luminous
interacting galaxies, HST has given us an impressive view of a myriad of Super
Star Clusters (SSCs) in their starburst regions. Even with ground based
observations, we have now been able to assess this complex structure,
particularly in the near-Infrared. The star forming knots in HII galaxies, as in
more luminous starbursts, are fragmented and consist of ensembles of star
clusters.  It seems, therefore, that the properties of the starburst are
independent from its cause, and when ignited it loses memory of its origin.

The temporal and spatial analysis of these star forming knots through a
combination of optical and near-IR observations have made it clear that the
history of star formation must be traced not only in time but in its spatial
distribution  \citep{tel02}. We have used the light gathering power and the
superb image quality of GEMINI in the near-IR (Lagos et al. 2009, in prep.) to
age date the resolved star clusters in a small sample of HII galaxies.
Figure~\ref{fig1} shows a spectacular example of our findings in the nearby HII
galaxy Mrk 36. Originally, this object was considered a bonafide example of a
compact young galaxy. As it can be seen, star clusters are numerous over
hundreds of parsec, with a regular envelope and no signs of double nucleus, nor
of being a product of a merger. 
young ages ($<$ a few $10^7$ yrs) and no sequential trend in ages across the
galaxy.  This alone rules out self-propagation as a dominant mechanism for star
formation in galactic scales.  These dwarf galaxies also lack spiral arms, so
density waves and shear are not considered important triggering mechanism. What
then triggers star formation in HII galaxies?

I wish to conclude this contribution by presenting a speculative view of the
mode of star formation in these regular, low luminosity, isolated HII galaxies.
The fact that the star clusters all have young ages and occur in galactic scales
($> 1$ Kpc) suggests that massive star cluster formation in starburst is
simultaneous within these time scales measured by age differences among the
SSCs. In other words, they occur in different places at the same time ($<$ a few
$10^7$ yrs) like ``pop corn'' in a hot oil pan! The ``oil'' in this caricature
picture is the interstellar medium and ``hot'' is the necessary physical
condition so that the ``corn'' (star clusters) will ``pop'' (form) here and
there in the ``pan'' (the observable starburst region).  What is then the
physical condition that governs the star formation magnitude and efficiency? 
This physical condition has to achieve the necessary threshold  simultaneously 
over these galactic scales so that star formation will occur in a stochastic
matter.  In this picture star cluster formation is simultaneous and stochastic
{\it within these time scales}.  This picture is all consistent and leads us
back to the work of \cite{sch63} who proposed the star formation rate (SFR) as a
power law relation with gas densities.  \cite{ken98} calibrated this law, in its
observable form (known as Kennicutt-Schmidt law), as a power law relation 
between the surface SFR and surface total gas (atomic and molecular) density for
star-forming disks of spirals, and extended to starburst galaxies. The recent
work from the THINGS collaboration \citep[see e.g.][]{ler08} has identified the
molecular gas densities to be primarily responsible for the proportionality with
SFR which seems to be a reasonable result since massive star and cluster
formation ultimately occur in giant molecular clouds. Therefore, once the
``oil'' (molecular gas)  is ``hot'' (reached a threshold surface density through
gravitational collapse and infall) enough, small pertubations \citep{ick85} will
trigger the ``corn'' (star clusters) to ``pop'' (form) here and there over the
``pan`` (extent of starburst).  Finally, the interplay of stellar feedback and
gas dynamics will lead to the duration and fate of the present starburst, before
it ceases and the galaxy undergoes a quiescent phase till gas falls back to heat
the oil in an episodic matter \citep{pel04}.

\acknowledgements 
{\small
ET acknowledges his US Gemini Fellowship by
AURA. Thanks to Evan Skillman, Zhi-Yun Li, Bob O'Connell and Trinh Thuan for reading this manuscript.}


\end{document}